\documentclass[fleqn,usenatbib,useAMS]{mnras}
 \usepackage{graphicx}
 \usepackage{amsmath}
 \usepackage{amssymb}
 \usepackage[T1]{fontenc}
 \usepackage{ae,aecompl}
 \usepackage{txfonts}

 \title[Impact of the Planet Nine hypothesis]
       {Dynamical impact of the Planet Nine scenario: \textit{N}-body experiments}

 \author[C. de la Fuente Marcos, R. de la Fuente Marcos and S. J. Aarseth]
        {Carlos~de~la~Fuente~Marcos,$^{1}$\thanks{E-mail: carlosdlfmarcos@gmail.com}
         Ra\'ul~de~la~Fuente Marcos$^{1}$
         and
         Sverre J. Aarseth$^2$ \\
         $^1$Apartado de Correos 3413, E-28080 Madrid, Spain \\
         $^2$Institute of Astronomy, University of Cambridge,
             Madingley Road, Cambridge CB3 0HA, UK}
 \date{Accepted 2016 April 21.
       Received 2016 April 21;
       in original form 2016 March 7}
 \pubyear{2016}
 
\begin{document}
  \label{firstpage}
  \pagerange{\pageref{firstpage}--\pageref{lastpage}}
  \maketitle

  \begin{abstract}
     The Planet Nine hypothesis has now enough constraints to deserve further 
     attention in the form of detailed numerical experiments. The results of 
     such studies can help us improve our understanding of the dynamical 
     effects of such a hypothetical object on the extreme trans-Neptunian 
     objects or ETNOs and perhaps provide additional constraints on the orbit 
     of Planet Nine itself. Here, we present the results of direct $N$-body 
     calculations including the latest data available on the Planet Nine 
     conjecture. The present-day orbits of the six ETNOs originally linked to 
     the hypothesis are evolved backwards in time and into the future under 
     some plausible incarnations of the hypothesis to investigate if the 
     values of several orbital elements, including the argument of 
     perihelion, remain confined to relatively narrow ranges. We find that a 
     nominal Planet Nine can keep the orbits of (90377) Sedna and 
     2012~VP$_{113}$ relatively well confined in orbital parameter space for 
     hundreds of Myr, but it may make the orbits of 2004~VN$_{112}$, 
     2007~TG$_{422}$ and 2013~RF$_{98}$ very unstable on time-scales of 
     dozens of Myr, turning them retrograde and eventually triggering their 
     ejection from the Solar system. Far more stable orbital evolution is 
     found with slightly modified orbits for Planet Nine.
  \end{abstract}


  \section{Introduction}
     The discovery of 2012~VP$_{113}$ (Trujillo \& Sheppard 2014) has led the astronomical community to recognize that most, if not all,  
     asteroids with semimajor axis, $a$, greater than 150 au and perihelion distance, $q=a(1-e)$, greater than 30 au ---the extreme 
     trans-Neptunian objects or ETNOs--- display unusual patterns in the values of some of their orbital elements. The values of their 
     argument of perihelion, $\omega$, cluster about 0\degr ---actually, $-$26\degr--- (de la Fuente Marcos \& de la Fuente Marcos 2014; 
     Trujillo \& Sheppard 2014), their longitude of the ascending node, $\Omega$, about 134\degr (Batygin \& Brown 2016; Brown \& Firth 
     2016), their eccentricities, $e$, about 0.8, and their inclinations, $i$, about 20\degr (de la Fuente Marcos \& de la Fuente Marcos 
     2014, 2016; de la Fuente Marcos, de la Fuente Marcos \& Aarseth 2015). The clustering in $e$ can be explained as resulting from a 
     selection effect (if $a$=150~au and $q$=30~au, then $e$=0.8), but this is not the case of the ones observed in $i$, $\Omega$ and 
     $\omega$ (de la Fuente Marcos \& de la Fuente Marcos 2014). One of the theories proposed to explain the observed patterns places their 
     sources in the perturbations due to yet undetected trans-Plutonian planets.

     Perhaps the most popular incarnation of the trans-Plutonian massive perturber(s) paradigm is the so-called Planet Nine hypothesis. 
     Based on results from extensive computer simulations, Batygin \& Brown (2016) have predicted the existence of a massive planet located 
     well beyond Pluto in order to explain the observed clustering in physical space of the perihelia of six ETNOs. The putative object 
     responsible for inducing the clumping has been provisionally designated Planet Nine. Batygin \& Brown (2016) have provided tentative 
     values for the orbital parameters of the proposed 10 $M_{\oplus}$ planet: $a=700$~au, $e=0.6$, $i=30$\degr and $\omega=150$\degr. The 
     resonant scenario described in Batygin \& Brown (2016) assumes that the orbits of the ETNOs are apsidally anti-aligned and nodally 
     aligned with that of the perturber. The extrapolation of the {\it Cassini} data carried out by Fienga et al. (2016) indicates that 
     Planet Nine as characterised by Batygin \& Brown (2016) cannot exist in the interval of true anomaly $f\in(-132, 106.5)$\degr. In 
     addition, Fienga et al. (2016) pointed out that from the perspective of the {\it Cassini} residuals, the most probable position of 
     Planet Nine ---assuming that $\Omega=113$\degr--- is at $f=117\fdg8^{+11\degr}_{-10\degr}$. In the following, we define a nominal 
     orbit of Planet Nine as the set of orbital elements: $a=700$~au, $e=0.6$, $i=30$\degr, $\Omega=113$\degr and $\omega=150$\degr, with 
     $f=117\fdg8$ at $t=0$ (present time, see below). Regarding its origin, Planet Nine may have been scattered out of the region of the 
     giant planets early in the history of the Solar system (see e.g. Bromley \& Kenyon 2014, 2016) or even captured from another planetary 
     system (Li \& Adams 2016; Mustill, Raymond \& Davies 2016). 

     Table \ref{6etnos} shows the values of various orbital parameters for the six objects discussed in Batygin \& Brown (2016) as well as 
     relevant descriptive statistics; in this table, unphysical values are displayed for completeness. Both 2007 TG$_{422}$ and 2012 
     VP$_{113}$ are statistical outliers in terms of $e$ as their values are outside the range (Q$_{1}-1.5$IQR, Q$_{3}+1.5$IQR), where 
     Q$_{1}$ is the first quartile, Q$_{3}$ is the third quartile and IQR is the interquartile range or difference between them; we will 
     see that, within the context of the Planet Nine hypothesis, 2007 TG$_{422}$ is particularly interesting. The data show that the mean 
     value of the aphelion distance, $Q=a(1+e)$, is 700~au and the lowest value of $a$ is 259~au. An orbit with $a=700$~au and $q=259$~au 
     has $e=0.63$. On the other hand, the highest value of $i$ is 30\degr and the mean values of $\Omega$ and $\omega$ are 102\degr and 
     314\degr, respectively. The actual values of these two angular orbital elements are critical within the context of a nodal alignment 
     and apsidal anti-alignment scenario. If the relative value of $\Omega$ should be 0\degr and the relative value of the longitude of 
     perihelion, $\varpi=\Omega+\omega$, is expected to be equal to 180\degr (see e.g. Klafke, Ferraz-Mello \& Michtchenko 1992), then the 
     values of $\Omega$ and $\omega$ for Planet Nine must be close to 102\degr and 134\degr, respectively. The combination of values of 
     $\Omega$ and $\omega$ may have a major impact on the overall stability of the ETNOs under Planet Nine's gravitational perturbation, and 
     it is therefore legitimate to question the plausibility of the values proposed so far.
%
%
      \begin{table*}
        \centering
        \fontsize{8}{11pt}\selectfont
        \tabcolsep 0.10truecm
        \caption{Values of various orbital parameters ---$P$ is the orbital period, $\Omega^*$ and $\omega^*$ are $\Omega$ and $\omega$ in 
                 the interval ($-{\rm \pi}$, ${\rm \pi}$) instead of the regular (0, 2${\rm \pi}$)--- for the six objects discussed in 
                 Batygin \& Brown (2016). The statistical parameters are Q$_{1}$, first quartile, Q$_{3}$, third quartile, IQR, 
                 interquartile range, OL, lower outlier limit (Q$_{1}-1.5$IQR), and OU, upper outlier limit (Q$_{3}+1.5$IQR); see the text 
                 for additional details. (Epoch: 2457400.5, 2016-January-13.0 00:00:00.0 TDB. J2000.0 ecliptic and equinox. Source: JPL 
                 Small-Body Database. Data retrieved on 2016 February 27.)
                }
        \begin{tabular}{lrrrrrrrrrrr}
          \hline
             Object             & $a$ (au) & $e$      & $i$ (\degr) & $\Omega$ (\degr) & $\omega$ (\degr) & $\varpi$ (\degr) & $q$ (au) & 
                       $Q$ (au) & $P$ (yr)  & $\Omega^*$ (\degr) & $\omega^*$ (\degr) \\
          \hline
        (90377) Sedna           & 507.5603 & 0.8501824 & 11.92872   & 144.5463         & 311.4614         &  96.0077         & 76.0415  &
                       939.0792 & 11435.094 &  144.5463          &  $-$48.5386        \\
                2004 VN$_{112}$ & 321.0199 & 0.8525664 & 25.56295   &  66.0107         & 327.1707         &  33.1814         & 47.3291  &
                       594.7106 &  5751.830 &   66.0107          &  $-$32.8293        \\
                2007 TG$_{422}$ & 492.7277 & 0.9277916 & 18.58697   & 112.9515         & 285.7968         &  38.7483         & 35.5791  &
                       949.8764 & 10937.517 &  112.9515          &  $-$74.2032        \\
                2010 GB$_{174}$ & 371.1183 & 0.8687090 & 21.53812   & 130.6119         & 347.8124         & 118.4243         & 48.7245  &
                       693.5121 &  7149.518 &  130.6119          &  $-$12.1876        \\
                2012 VP$_{113}$ & 259.3002 & 0.6896024 & 24.04680   &  90.8179         & 293.7168         &  24.5346         & 80.4862  &
                       438.1142 &  4175.538 &   90.8179          &  $-$66.2832        \\
                2013 RF$_{98}$  & 309.0738 & 0.8826022 & 29.61402   &  67.5205         & 316.4991         &  24.0196         & 36.2846  &
                       581.8631 &  5433.774 &   67.5205          &  $-$43.5009        \\
          \hline
             Mean               & 376.8000 & 0.8452423 & 21.87960   & 102.0764         & 313.7429         &  55.8193         & 54.0742  &
                       699.5259 &  7480.545 &  102.0764          &  $-$46.2571        \\ 
             Standard deviation & 102.0532 & 0.0813175 &  6.13291   &  32.7348         &  22.5205         &  40.8126         & 19.5592  &
                       206.5256 &  3026.532 &   32.7348          &   22.5205          \\
             Median             & 346.0691 & 0.8606377 & 22.79246   & 101.8847         & 313.9803         &  35.9648         & 48.0268  &
                       644.1114 &  6450.674 &  101.8847          &  $-$46.0197        \\
             Q$_{1}$            & 312.0603 & 0.8507784 & 19.32476 &  73.3448           & 298.1529         &  26.6963         & 39.0457  &
                       585.0750 &  5513.288 &   73.3448          &  $-$61.8471        \\
             Q$_{3}$            & 462.3254 & 0.8791289 & 25.18391 & 126.1968           & 324.5028         &  81.6929         & 69.2122  &
                       877.6874 &  9990.517 &  126.1968          &  $-$35.4972        \\
             IQR                & 150.2650 & 0.0283505 &  5.85915 &  52.8520           &  26.3499         &  54.9966         & 30.1665  &
                       292.6125 &  4477.229 &   52.8520          &   26.3499          \\
             OL                 & 86.6628  & 0.8082526 & 10.53603 & $-$5.9331          & 258.6281         & $-$55.7985       & $-$6.2041&
                       146.1563 & $-$1202.556 & $-$5.9331        & $-$101.3719        \\
             OU                 & 687.7230 & 0.9216547 & 33.97264 & 205.4747           & 364.0277         & 164.1877         & 114.4620 &
                      1316.6061 & 16706.361 &  205.4747          &    4.0277          \\
          \hline
        \end{tabular}
        \label{6etnos}
      \end{table*}
%
%

     Assessing the impact of the gravitational perturbation from a Planet Nine-like perturber on the short-term evolution of the known ETNOs 
     is essential in order to decide if a candidate orbit is plausible or not. Here, we present the results of direct $N$-body calculations 
     including the latest data available on the Planet Nine conjecture. The results from these calculations can help us improve our 
     understanding of the dynamical effects of such a hypothetical object, or any other trans-Plutonian perturber, on the ETNO population. 
     This Letter is organized as follows. Our $N$-body methodology is briefly described in Section 2. The evolution of the six ETNOs 
     included in Table \ref{6etnos} subjected to the perturbation of Planet Nine moving along the nominal orbit described in Fienga et al. 
     (2016) is studied in Section 3. Section 4 repeats the analysis for other orbits. Results are discussed and conclusions summarized in 
     Section~5. 

  \section{Assessing the dynamical effects: an \textit{N}-body approach}
     It is a well-known fact that the motion of the planets in the Solar system is chaotic; accurate predictions of their trajectories 
     beyond a few tens of Myr are simply not possible (see e.g. Laskar 1989, 1990, 1994, 2008). Extensive computer simulations show that the 
     Solar system as we know it remains at all times in a state of marginal stability; it is therefore reasonable to assume that the 
     inclusion of Planet Nine must not change this current state of dynamical affairs. On the other hand, the known planets are not 
     distributed randomly but organize themselves into planetary groups or collections of loosely connected mutually dynamically dependent 
     planets. Innanen, Mikkola \& Wiegert (1998) and Ito \& Tanikawa (1999, 2002) have shown numerically that the terrestrial planets 
     maintain their stability by sharing and weakening the secular perturbation from Jupiter. Tanikawa \& Ito (2007) further extended this 
     analysis concluding that the outer planetary group, or Jovian planets, forms a subsystem that is not affected by the inner planets; the 
     motion of the group of the Jovian planets may not be perturbed to a significant degree if there is no inner planet group. This 
     numerical evidence can be used to set up the most efficient yet still reliable physical model to perform $N$-body experiments that 
     include an implementation of the Planet Nine hypothesis for testing. In other words, any realistic $N$-body study of the dynamical 
     effects of the Planet Nine conjecture must include the Jovian planets, all of them, but can safely neglect the terrestrial planets. 
     Under such physical model, we should expect accurate results within a few tens of Myr of $t=0$; if a virtual ETNO becomes dynamically 
     unstable in a few tens of Myr, this can be considered as a robust feature of that particular simulation and the same can be said about 
     an orbit that remains fairly stable for a similar period of time.

\begin{figure*}
        \centering
         \includegraphics[width=0.24\linewidth]{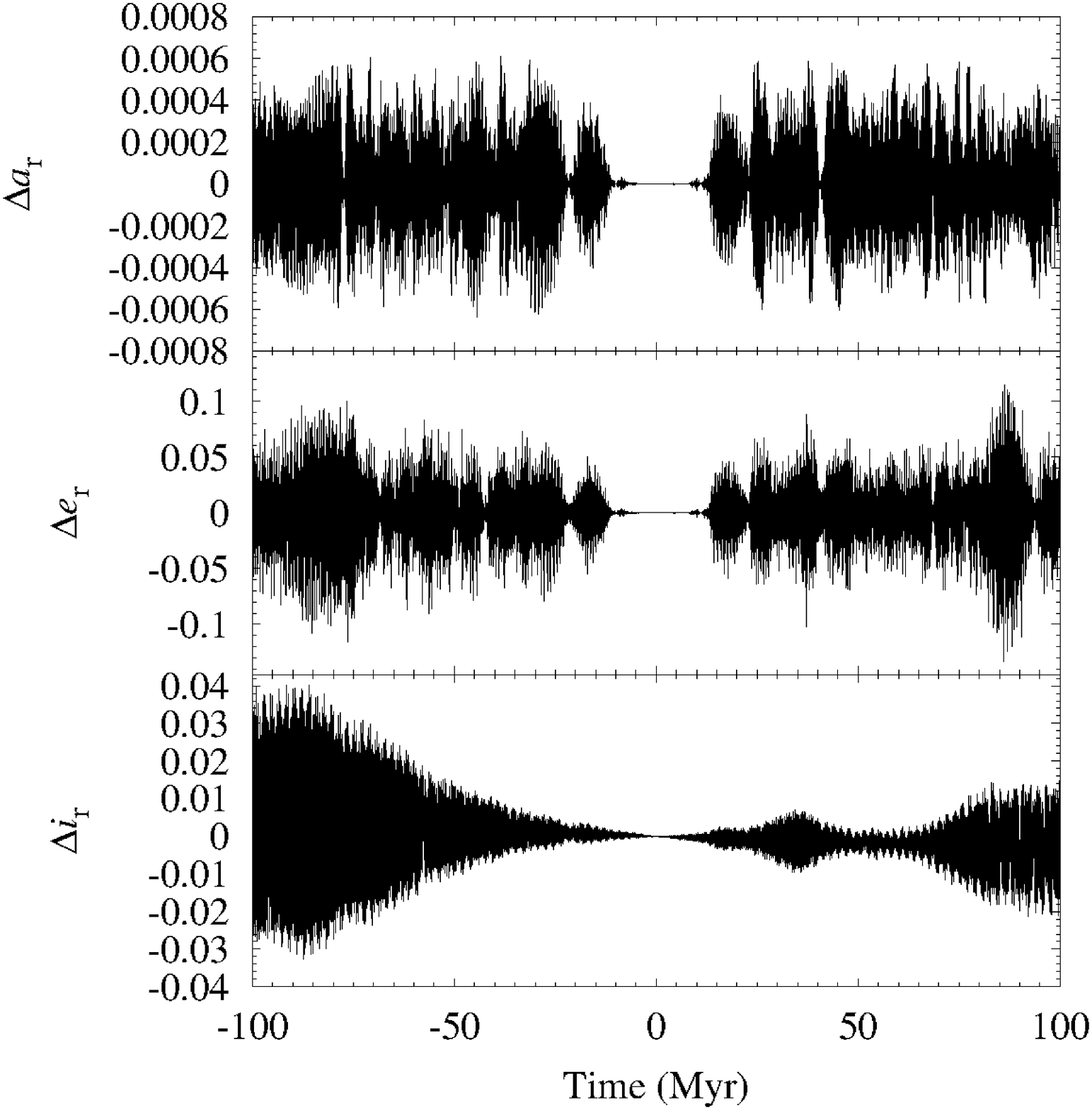}
         \includegraphics[width=0.24\linewidth]{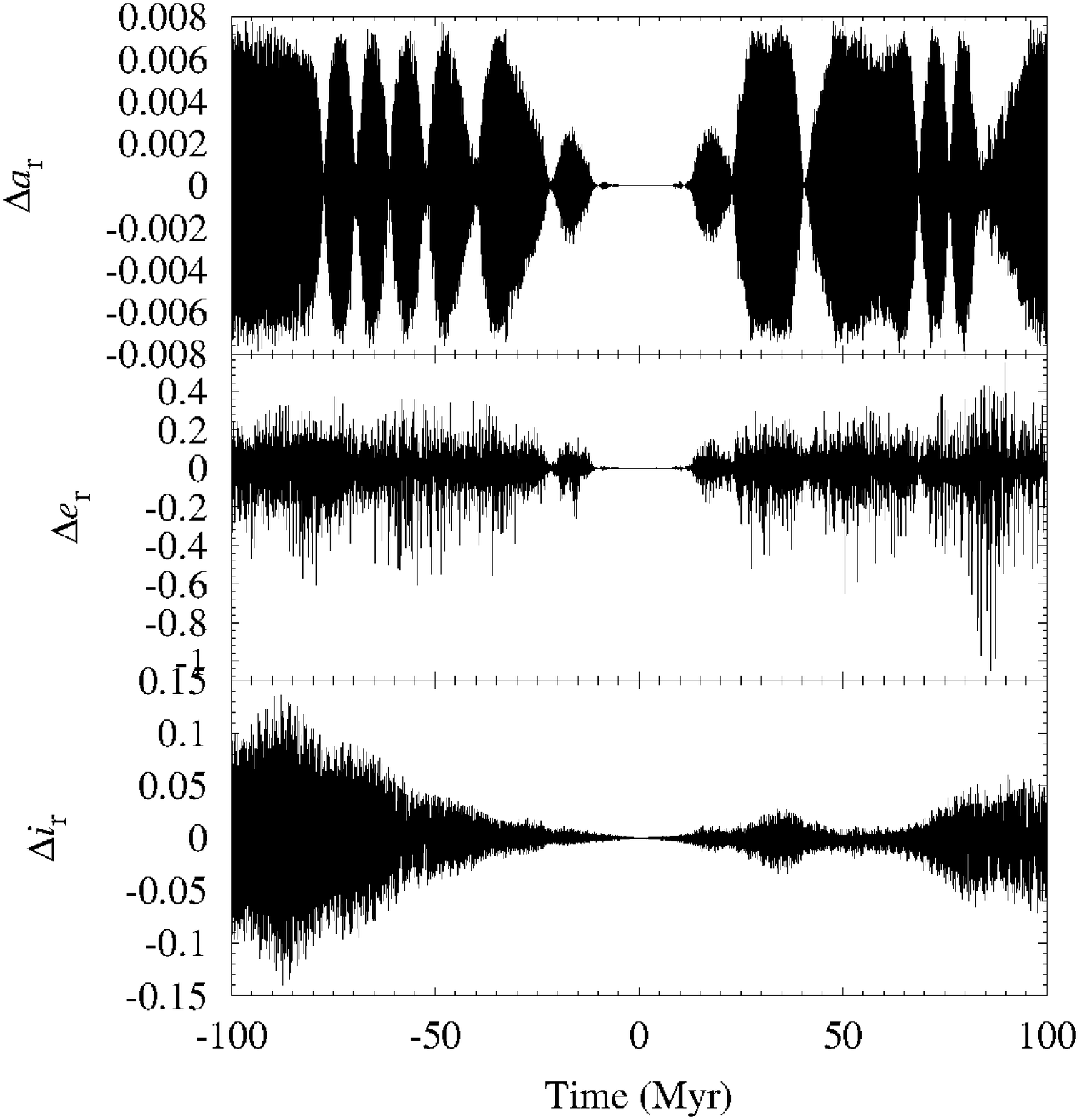}
         \includegraphics[width=0.24\linewidth]{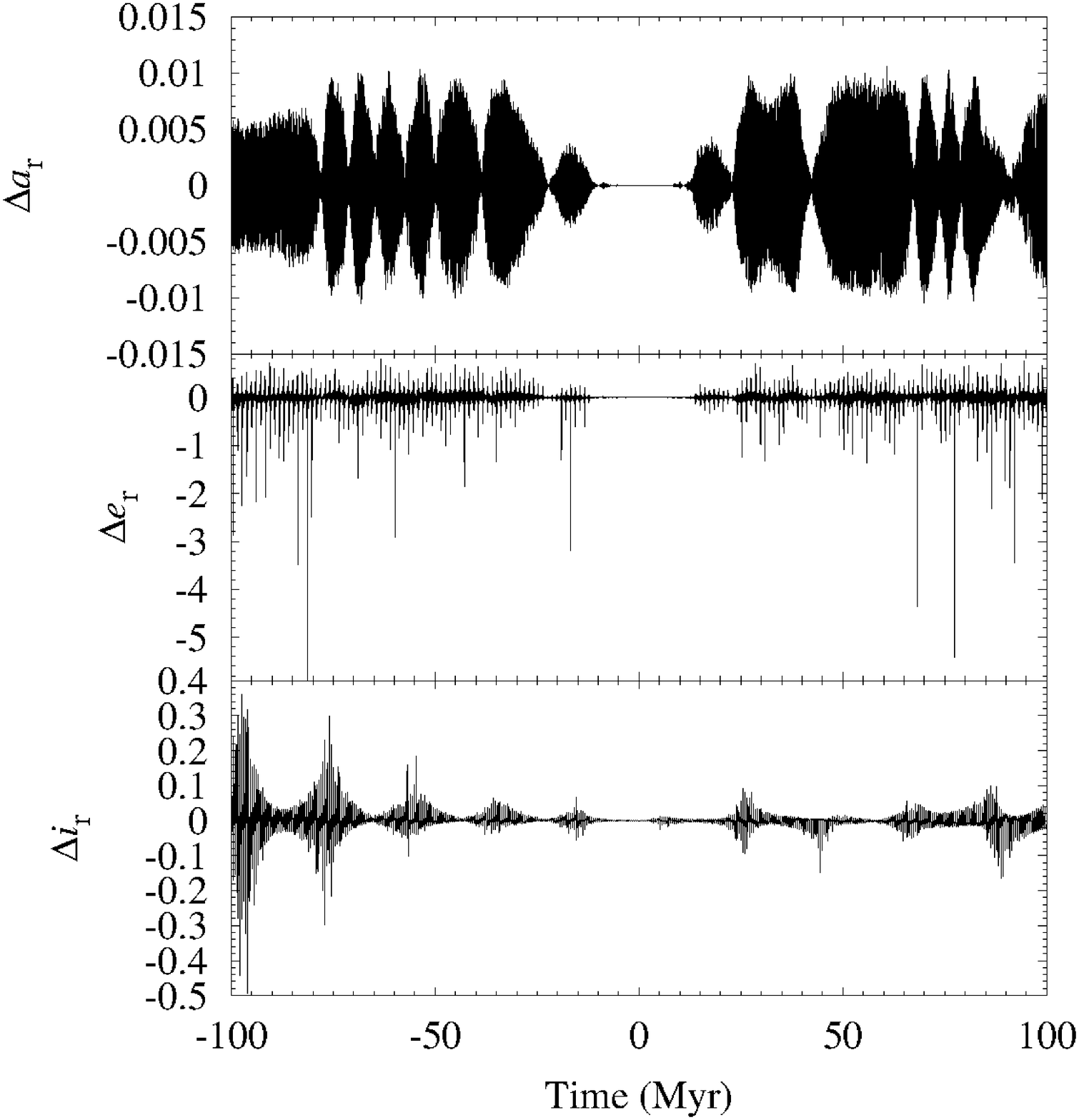}
         \includegraphics[width=0.24\linewidth]{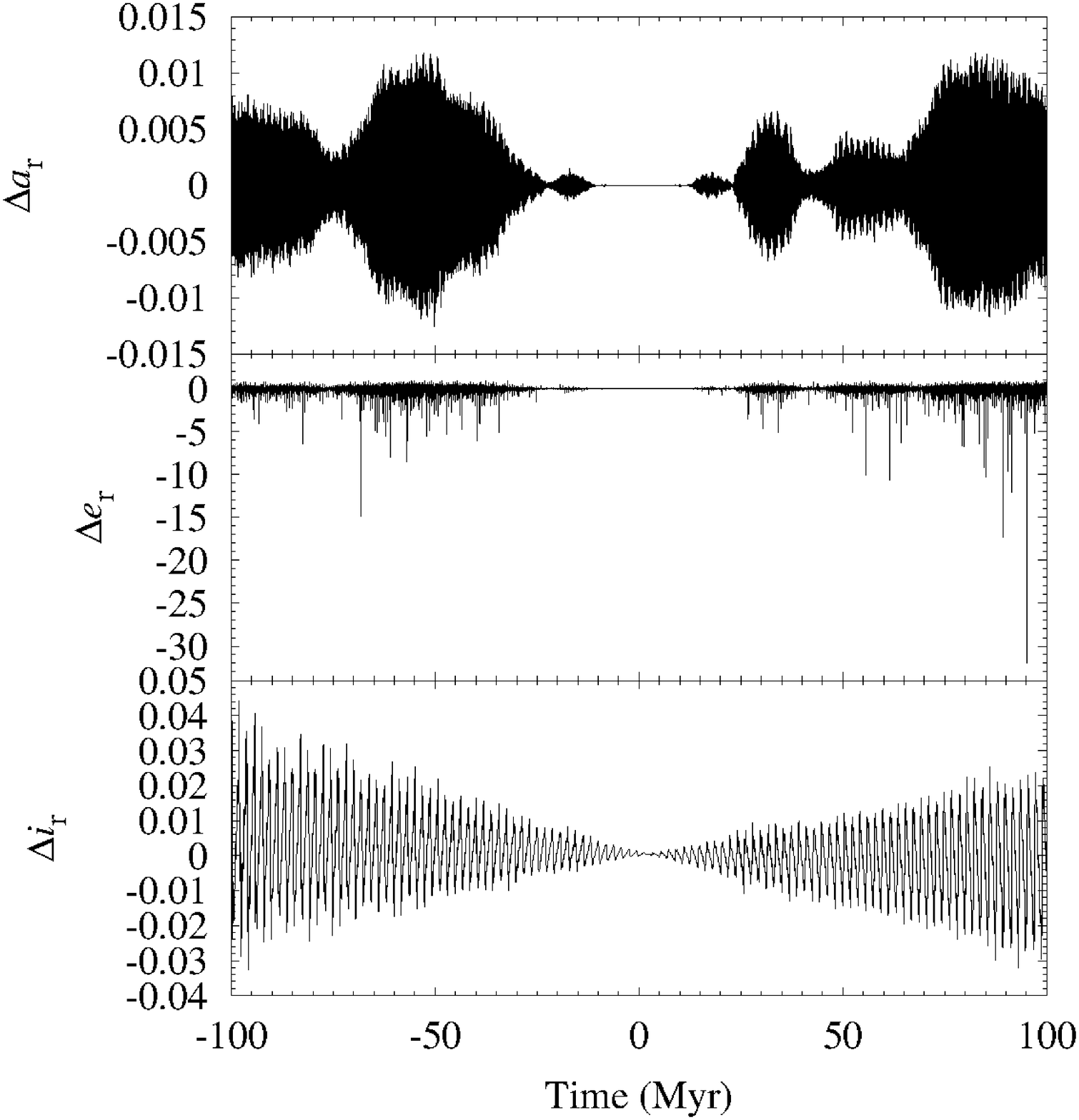} \\
         \caption{Relative variations, with respect to the case without Planet Nine, of the values of the orbital parameters semimajor axis, 
                  eccentricity and inclination of the Jovian planets ---Jupiter, left-hand panels, to Neptune, right-hand panels--- 
                  subjected to the perturbation due to the nominal orbit of Planet Nine. 
                 }
         \label{deviations}
      \end{figure*}

     Consistently with the above discussion, our physical model includes the perturbations by the Jovian planets (Jupiter to Neptune). In 
     order to compute accurate initial positions and velocities we used the heliocentric ecliptic Keplerian elements provided by the JPL 
     On-line Solar System Data Service\footnote{http://ssd.jpl.nasa.gov/?planet\_pos} (Giorgini et al. 1996) and initial positions and 
     velocities (for both planets and ETNOs) based on the DE405 planetary orbital ephemerides (Standish 1998) referred to the barycentre of 
     the Solar system and to the epoch JD TDB 2457400.5 (2016-January-13.0), which is the $t$ = 0 instant in our figures. In addition to the 
     orbital calculations completed using the nominal elements of the ETNOs in Table \ref{6etnos}, we have performed additional simulations 
     using control orbits; the orbital elements of the control orbits have been obtained varying them randomly, within the ranges defined by 
     their mean values and standard deviations (1$\sigma$). In our calculations, the Hermite integration scheme described by Makino (1991) 
     and implemented by Aarseth (2003) is employed. The standard version of this direct $N$-body code is publicly available from the IoA 
     website.\footnote{http://www.ast.cam.ac.uk/$\sim$sverre/web/pages/nbody.htm} Relative errors in the total energy for the longest 
     integrations are as low as $3\times10^{-12}$ or lower. The relative error in the total angular momentum is several orders of magnitude 
     smaller. As pointed out in de la Fuente Marcos \& de la Fuente Marcos (2012), the results from this code compare well with those from 
     Laskar et al. (2011) among others. Fig. \ref{deviations} shows the relative variations (difference between the values with and 
     without Planet Nine divided by the value without the external perturber) of the values of the orbital elements ---$a$, $e$, $i$--- of 
     the four Jovian planets (Jupiter to Neptune, left to right) for $\pm$100 Myr for the nominal orbit of Planet Nine. These deviations 
     are consistent with the expectations as they remain very small for a few tens of Myr, growing afterwards. The average orbital evolution 
     of the Jovian planets, which is time symmetric, is not significantly altered by the presence of Planet Nine as described by the nominal 
     orbit (see above).  
%
%
%
%

     Batygin \& Brown (2016) used the {\sc mercury}6 gravitational dynamics software package (Chambers 1999) to perform $N$-body 
     simulations of 13 ETNOs subjected to the perturbation of the most massive known members of the Solar system. These calculations led 
     them to select six out of 13 ETNOs as largely unaffected by the presence of Neptune; these objects are (90377) Sedna, 2004~VN$_{112}$, 
     2007~TG$_{422}$, 2010~GB$_{174}$, 2012~VP$_{113}$ and 2013~RF$_{98}$. The present-day orbital properties of these objects were 
     subsequently used to constrain and validate the resonant perturbation mechanism behind the Planet Nine hypothesis. However, no attempt 
     was made to confirm that, in the presence of their favoured incarnation of the Planet Nine hypothesis, the orbital evolution of the six 
     ETNOs remained unaffected by Neptune. In the following, we perform such tests ---evolving the present-day orbits of the six ETNOs 
     backwards in time and into the future--- within the framework described above and discuss the results obtained.

  \section{Impact of the orbit in Fienga et al. (2016)}
\begin{figure}
        \centering
         \includegraphics[width=\linewidth]{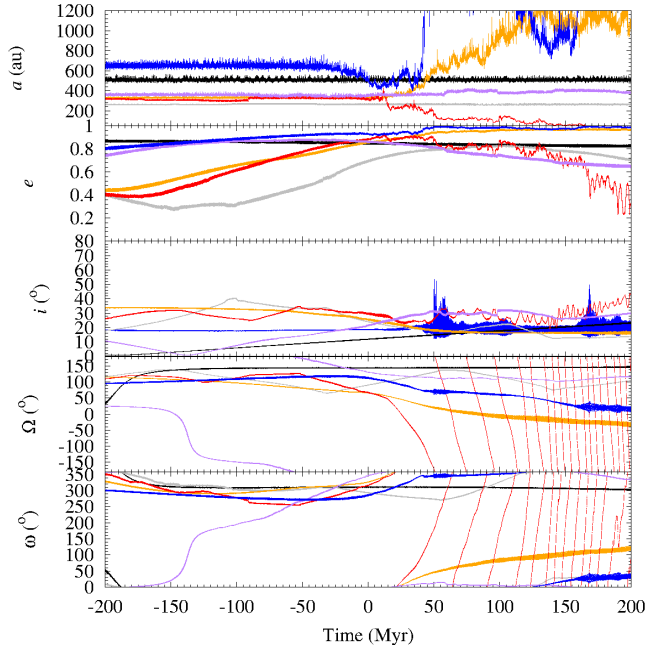}
         \caption{Orbital evolution of the six ETNOs in Table \ref{6etnos} subjected to the nominal Planet Nine perturbation (see the text 
                  for details): (90377) Sedna (black), 2004~VN$_{112}$ (orange), 2007~TG$_{422}$ (blue), 2010~GB$_{174}$ (purple), 
                  2012~VP$_{113}$ (grey) and 2013~RF$_{98}$ (red). 
                 }
         \label{fienga}
      \end{figure}

     Fienga et al. (2016) pointed out that from the perspective of the {\it Cassini} residuals, the most probable orbit for the object 
     proposed by Batygin \& Brown (2016) could be: $a=700$~au, $e=0.6$, $i=30$\degr, $\Omega=113$\degr and $\omega=150$\degr, with 
     $f=117\fdg8$ at present time. We have computed the orbital evolution of the six ETNOs included in Table \ref{6etnos} subjected to the 
     perturbation of Planet Nine for $\pm200$~Myr. Our results are summarized in Fig. \ref{fienga} that corresponds to the nominal orbits of 
     the six ETNOs subjected to the action of planet Nine as discussed in Fienga et al. (2016). They indicate that the orbital evolution of 
     several of the objects displayed could be more chaotic into the future than it used to be in the past. This time asymmetry suggests 
     that, within the Planet Nine hypothesis, the ETNOs might be a transient population that evolves from relatively stable orbits into 
     rather unstable ones. In particular, the future orbital evolution of 2004~VN$_{112}$, 2007~TG$_{422}$ and 2013~RF$_{98}$ is very 
     unstable in this case. Using control orbits for the ETNOs slightly different from the nominal ones, but within 1$\sigma$, produces 
     similar results. The same can be said if it is assumed that Planet Nine is currently located at aphelion and/or that $\Omega\sim90$\degr 
     but leaving the values of the other orbital parameters unchanged.
%
%
%
%
%
%
      \begin{figure}
        \centering
         \includegraphics[width=\linewidth]{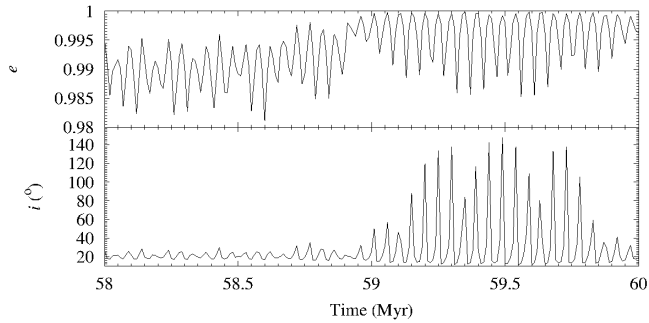}
         \caption{Detail of the pre-ejection evolution of 2007~TG$_{422}$ for a simulation different from the one displayed in 
                  Fig. \ref{fienga}. 
                 }
         \label{retro}
      \end{figure}
%
%
     Fig. \ref{fienga} clearly shows that the orbital evolution of several of the six ETNOs is eventually affected by Neptune (and other
     Jovian planets) as their eccentricities become close to 1 (2004~VN$_{112}$ and 2007~TG$_{422}$) or its semimajor axis becomes 
     dangerously small (2013~RF$_{98}$). Fig. \ref{retro} shows the pre-ejection evolution of one of the control orbits of 2007~TG$_{422}$
     that ends in ejection from the Solar system; the object turns retrograde following a dynamical pathway that resembles closely the one 
     described for comet 96P/Machholz 1 in de la Fuente Marcos et al. (2015). Multiple orbit realizations of the ETNOs 2004~VN$_{112}$, 
     2007~TG$_{422}$ and 2013~RF$_{98}$ evolve into retrograde objects and/or are ejected from the Solar system within the simulation time.
     This behaviour has been previously documented in simulations by Gomes, Soares \& Brasser (2015) and Batygin \& Brown (2016). Only the 
     orbits of (90377) Sedna and 2012~VP$_{113}$ remain relatively well confined in orbital parameter space in Fig. \ref{fienga}. However, 
     longer integrations (not shown) indicate that, for this particular orbit, the six ETNOs could be ejected from the Solar system within 
     1.5 Gyr from now. It is clear that, in general, the assumption made by Batygin \& Brown (2016) ---that the singled out ETNOs were 
     largely unaffected by the presence of Neptune--- breaks up relatively rapidly once the perturbation from Planet Nine is taken into 
     account. This is to be expected within the context of an apsidal anti-alignment scenario because the average value of $\Delta\varpi$ 
     ---using this hypothetical orbit and the data from Table \ref{6etnos}--- is nearly 153\degr, relatively far from the critical value of 
     180\degr. So the question is, how is the orbital evolution of the ETNOs affected if we enforce the $\Delta\varpi\sim180$\degr 
     condition?

  \section{Impact of other solutions}
     Here, we present results from two representative simulations that consider candidate Planet Nine orbits with $\Delta\Omega$$\sim$0\degr 
     and $\Delta\varpi$$\sim$180\degr.

     \subsection{Orbit A}
\begin{figure}
        \centering
         \includegraphics[width=\linewidth]{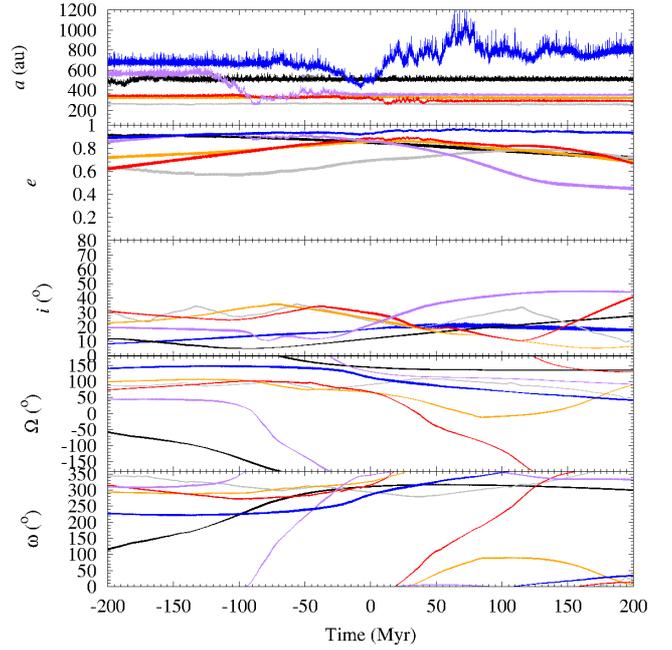}
         \caption{Similar to Fig. \ref{fienga} but for one representative solution with $\Delta\varpi\sim180$\degr (see the text for 
                  details). 
                 }
         \label{subaru}
      \end{figure}

        Fig. \ref{subaru} shows the orbital evolution of the same objects displayed in Fig. \ref{fienga} but subjected to the perturbation 
        of a 10 $M_{\oplus}$ planet moving along the orbit: $a=701$~au, $e=0.6$, $i=33$\degr, $\Omega=89$\degr and $\omega=142$\degr, with 
        $f=180$\degr at $t=0$. Now $\Delta\varpi=185$\degr and the overall evolution is far more stable than that in Fig. \ref{fienga}; no 
        objects turn retrograde and/or are ejected during the displayed time. The values of $e$ and $i$ remain within narrow limits, but 
        $\omega$ circulates for two objects, 2010~GB$_{174}$ and 2013~RF$_{98}$. The values of the orbital elements of 2007~TG$_{422}$, that 
        was the most unstable in Section 3, remain well confined. 
%
%
%
%
     \subsection{Orbit B}
\begin{figure}
        \centering
         \includegraphics[width=\linewidth]{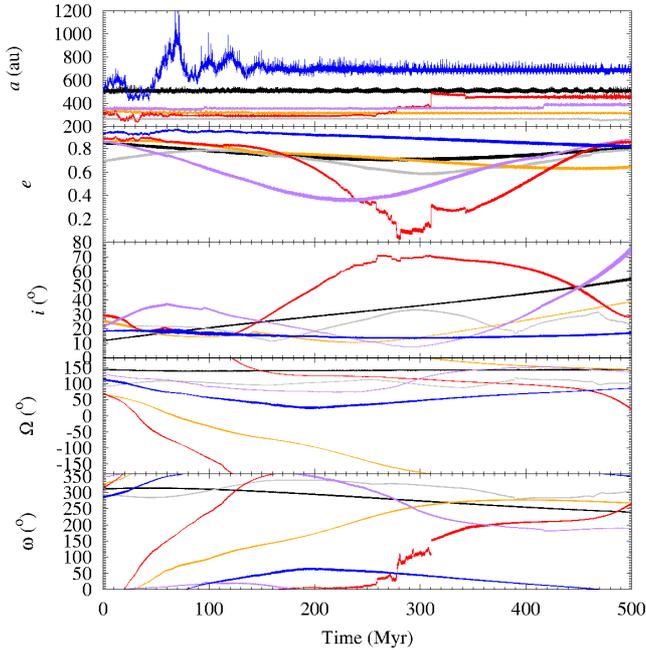}
         \caption{Orbital evolution into the future for a representative solution with $\Delta\Omega=0$\degr and $\Delta\varpi=180$\degr 
                  (see the text for details). Colours as in Fig. \ref{fienga}.
                 }
         \label{stable}
      \end{figure}

        The previous calculations suggest that the orbital evolution of the ETNOs subjected to Planet Nine perturbation could be more 
        unstable into the future. For this reason, in this integration we neglect the past evolution of the ETNOs and focus on the future. 
        The orbit of Planet Nine used here is: $a=700$~au, $e=0.6$, $i=30$\degr, $\Omega=102$\degr and $\omega=134$\degr, with $f=180$\degr 
        at $t=0$. In this case $\Delta\Omega=0$\degr and $\Delta\varpi=180$\degr. Fig. \ref{stable} shows the orbital evolution of the same 
        objects displayed in Figs \ref{fienga} and \ref{subaru}. Now, 2007~TG$_{422}$ becomes an eccentric but stable co-orbital trapped in 
        the 1:1 mean motion resonance with Planet Nine after $\sim$160 Myr (this is also observed sometimes for orbit A). Co-orbital motion 
        is possible at high eccentricity (see e.g. Namouni, Christou \& Murray 1999). The overall orbital evolution of the six ETNOs under 
        this incarnation of Planet Nine is more stable than that in Fig. \ref{fienga}, but less stable than the one associated with orbit A 
        (see Fig. \ref{subaru}). Several objects exhibit Kozai dynamics (Kozai 1962), including 2012~VP$_{113}$ that also experiences this 
        behaviour for the previous two orbits.
%
%
%
%

  \section{Discussion and conclusions}
     The resonant coupling mechanism described in Batygin \& Brown (2016), with some fine tuning, can shepherd some ETNOs and explain their
     unusual orbital patterns. It can also create dynamical pathways that may deliver objects to high inclination or even retrograde orbits, 
     and place ETNOs within Kozai resonances. This last feature is shared with the scenarios discussed in Trujillo \& Sheppard (2014), de la 
     Fuente Marcos \& de la Fuente Marcos (2014) or de la Fuente Marcos et al. (2015) although the origin is different. Within the Planet
     Nine paradigm, Kozai behaviour is a by-product of mean motion resonances. 

     Innanen et al. (1997) and Tanikawa \& Ito (2007) performed numerical experiments to measure the strength of the gravitational coupling 
     among the known planets against external perturbation. They concluded that, for relatively short integrations, the two subsystems 
     pointed out above were able to absorb efficiently a perturbation several orders of magnitude stronger than that of a putative Planet 
     Nine. This result can be used to guess the answer to the question, is Planet Nine alone or are there more? Planet Nine, if it exists, 
     moves in an elongated orbit that may be vulnerable to long-term secular perturbations resulting from the Galactic tide or discrete 
     events like close encounters with passing stars. In this context, a lone Planet Nine may not be able to survive in its present orbit 
     for the age of the Solar system (see Li \& Adams 2016), but a planet within a planetary group has better chances to be long-term 
     stable. Therefore, if Planet Nine exists, it is probably not alone; planets similar to Uranus or Neptune (super-Earths) may also form 
     at 125--750 au from the Sun (Kenyon \& Bromley 2015, 2016).

     In this Letter, we have explored the dynamical impact of the Planet Nine scenario on the six ETNOs originally linked to the hypothesis
     proposed by Batygin \& Brown (2016). This study has been performed using $N$-body techniques. Our main conclusions are as follows. 
     \begin{itemize}
        \item The nominal orbit of Planet Nine as described in Fienga et al. (2016) can keep the orbits of (90377) Sedna and 2012~VP$_{113}$ 
              relatively well confined in orbital parameter space for hundreds of Myr but it may make the orbits of 2004~VN$_{112}$, 
              2007~TG$_{422}$ and 2013~RF$_{98}$ very unstable on time-scales of dozens of Myr, turning them retrograde and triggering their 
              ejection from the Solar system.  
        \item Modifying the orbit in Fienga et al. (2016) by assuming that Planet Nine is currently located at aphelion and/or that 
              $\Omega\sim90$\degr\ does not make the orbits of 2004~VN$_{112}$, 2007~TG$_{422}$ and 2013~RF$_{98}$ more stable and 
              short-term ejections are still observed.
        \item Slightly modified solutions for Planet Nine's orbit that enforce apsidal anti-alignment produce a far more stable dynamical 
              evolution of 2004~VN$_{112}$, 2007~TG$_{422}$ and 2013~RF$_{98}$.
        \item Orbital solutions with $\Delta\Omega\sim0$\degr and $\Delta\varpi\sim180$\degr\ induce Kozai dynamical evolution for several 
              of the ETNOs studied here.
        \item The orbital evolution of 2007~TG$_{422}$ is particularly sensitive to the details of the Planet Nine hypothesis. In some
              implementations the object turns retrograde to be eventually ejected from the Solar system, but in others it becomes a stable 
              co-orbital of Planet Nine. 
        \item For the nominal Planet Nine hypothesis, the ETNOs are a transient population with a typical lifetime of a few hundreds of 
              Myr.
     \end{itemize}

  \section*{Acknowledgements}
     We thank the anonymous referee for his/her constructive and helpful report, and D. P. Whitmire, G. Carraro, D. Fabrycky, A. V. Tutukov, 
     S. Mashchenko, S. Deen and J. Higley for comments on ETNOs and trans-Plutonian planets. This work was partially supported by the 
     Spanish `Comunidad de Madrid' under grant CAM S2009/ESP-1496. In preparation of this Letter, we made use of the NASA Astrophysics Data 
     System, the ASTRO-PH e-print server and the MPC data server.

  \bsp
  \label{lastpage}
\end{document}